\documentclass[10pt]{article}

\usepackage[margin=1in]{geometry} % adjust if you want slightly smaller/larger margins
\usepackage[utf8]{inputenc}
\usepackage{amssymb}
\usepackage{bm}
\usepackage{amsmath}
\usepackage[ruled]{algorithm2e}
\usepackage[usenames,dvipsnames]{color}
\usepackage{array}
\usepackage{subcaption}
\usepackage{graphicx}
\usepackage{lineno}
\usepackage{tikz}
\usetikzlibrary{shapes, arrows.meta, positioning, calc}
\usepackage{booktabs}
\usepackage{float}
\usepackage[labelfont=bf,textfont=it]{caption}
\usepackage{placeins}
\usepackage{longtable}
\usepackage{enumitem}

\usepackage{xurl}      % better line breaking for URLs
\usepackage{hyperref}
\hypersetup{
    colorlinks,
    linkcolor={magenta},
    citecolor={blue},
    urlcolor={blue!80!black},
    breaklinks=true,
    plainpages=true
}

\usepackage[numbers,sort&compress]{natbib}

\usepackage{etoolbox}
\apptocmd{\thebibliography}{\raggedright}{}{}

\newcommand{\cref}[3]{\hyperref[#2]{#1~\ref*{#2}{#3}}}
\newcommand{\crefs}[3]{\hyperref[#2]{#1~\ref*{#2}-\ref*{#3}}}
\newcommand{\colref}[2]{\hyperref[#2]{#1~\ref*{#2}}}

\title{Towards smart canopies: Algorithmic design of maize canopy architectures that maximize light use efficiency}

\author{
    Nasla Saleem$^{1}$ \and
    Talukder Zaki Jubery$^{1}$ \and
    Yan Zhou$^{2}$ \and
    Yawei Li$^{2}$ \and
    Adarsh Krishnamurthy$^{1,3}$ \and
    Patrick S. Schnable$^{2,3}$\thanks{Corresponding authors.} \and
    Baskar Ganapathysubramanian$^{1,3}$\footnotemark[1]
}

\date{
    $^{1}$Department of Mechanical Engineering, Iowa State University\\
    $^{2}$Department of Agronomy, Iowa State University\\
    $^{3}$Plant Science Institute, Iowa State University\\[0.5em]
}

\begin{document}

\maketitle

\begin{abstract}

% Your main text starts here.

We present a computational framework that integrates functional–structural plant modeling (FSPM) with an evolutionary algorithm to optimize three-dimensional maize canopy architecture for enhanced light interception under high-density planting. The optimization revealed an emergent ideotype characterized by two distinct strategies: a vertically stratified leaf profile (steep, narrow upper leaves for penetration; broad, horizontal lower leaves for capture) and a radially tiled azimuthal arrangement that breaks the conventional distichous symmetry of maize to minimize self- and mutual shading. Reverse ray-tracing simulations show that this architecture intercepts significantly more photosynthetically active radiation (PAR) than virtual canopies parameterized from high-performing field hybrids, with gains that generalize across multiple U.S. latitudes and planting densities. The optimized trait combinations align with characteristics of modern density-tolerant cultivars, supporting biological plausibility. Because recent gene-editing advances enable more independent control of architectural traits, the designs identified here are increasingly feasible.
By uncovering effective, non-intuitive trait configurations, our approach provides a scalable, predictive tool to guide breeding targets, improve light-use efficiency, and ultimately support sustainable yield gains.
\end{abstract}


% \linenumbers

%%%%%%%%%%%%%%%%%%%%%%%%%%%%%%%%%%%%%%%%%%%%%%%%%%%%%%%%%%%%%
%% main text

\section{Introduction}

Maize (\textit{Zea mays L.}) is a cornerstone of global agriculture, with annual production exceeding 1.1 billion metric tons and serving as a primary source of food, feed, fuel, and industrial products~\cite{erenstein2022global, nyirenda2021delving}. As the world’s population continues to grow and climate pressures intensify, maximizing maize yields on existing cropland is paramount. Increasing agricultural production on the same land base also reduces pressure to convert natural ecosystems, such as the Amazon rainforest, into  agricultural production, thereby helping to preserve biodiversity and carbon-rich landscapes. Over the past century, one of the most successful strategies for yield improvement has been increasing planting density: in the United States, average planting densities rose from roughly 30,000 plants\,ha$^{-1}$ in the mid-20th century to over 80,000 plants\,ha$^{-1}$ today~\cite{assefa2018analysis, yan2024photosynthetic, kalogeropoulos2024historical, duvick2005contribution}. This intensification was enabled by selecting at high plant densities for higher-yielding hybrids. Selection for higher yield under these conditions was associated with indirect selection of hybrids with more upright leaves and compact canopies that mitigate self-shading and improve light capture in dense stands. Yet, further gains via increased density alone are increasingly constrained by inter-plant competition and mutual shading, which deprive lower canopy leaves of light and cap whole-canopy photosynthesis~\cite{mansfield2014survey, saleem2025accessing}.

The concept of an idealized crop architecture, or ideotype, was first articulated by Donald in 1968 as a set of trait values optimized for a given environment and management system~\cite{donald1968breeding, mock1975ideotype, carbajal2024role, gauffreteau2018using, sedgley1991appraisal}. In grain crops, the so-called “smart canopy” ideotype balances light distribution such that top leaves are erect, allowing light to penetrate, while lower leaves are more horizontal to intercept the attenuated irradiance deep in the canopy~\cite{ort2015redesigning}. In maize, studies have borne out this paradigm: genotypes exhibiting a vertical gradient of leaf angles demonstrate enhanced light penetration and yield under high densities~\cite{jiang2025leaf, pepper1977leaf, han2024molecular}. Breeding for higher yield over the past several decades has also increased the prevalence of erect upper leaves, a trait consistently associated with improved light distribution and density tolerance and noted as a contributing factor to historical yield gains~\cite{duvick2005contribution}. However, recent analyses indicate that the rate of genetic gain in vertical leaf angle itself is slowing~\cite{elli2023maize}, suggesting that further improvement of this trait may be reaching its practical limits under current agronomic planting densities. This plateau underscores the need to identify and optimize additional canopy traits beyond upright leaves at the top to further improve light distribution and sustain yield gains in dense plantings.

% For example, the recently characterized \texttt{lac1} mutant, exhibiting upright upper leaves and relatively flat lower leaves, achieved roughly a 10–25\% increase in grain yield over field hybrids in very high dense plantings, underscoring that purposeful reconfiguration of leaf architecture can translate directly into productivity gains~\cite{tian2024maize}.
%Traditional breeding, however, has relied on incremental, heuristic approaches which are guided by breeders’ intuition (breeder's eye) and expensive, time-consuming field trials to accumulate favorable architectural traits. The sheer dimensionality of canopy design-leaf length, width, vertical leaf angles, azimuthal leaf angles, plant height, and beyond—renders exhaustive empirical exploration infeasible. 

\begin{figure*}[t!]
 \centering
 \includegraphics[width=\linewidth, height=0.77\textheight, trim=0 80 0 0, clip]{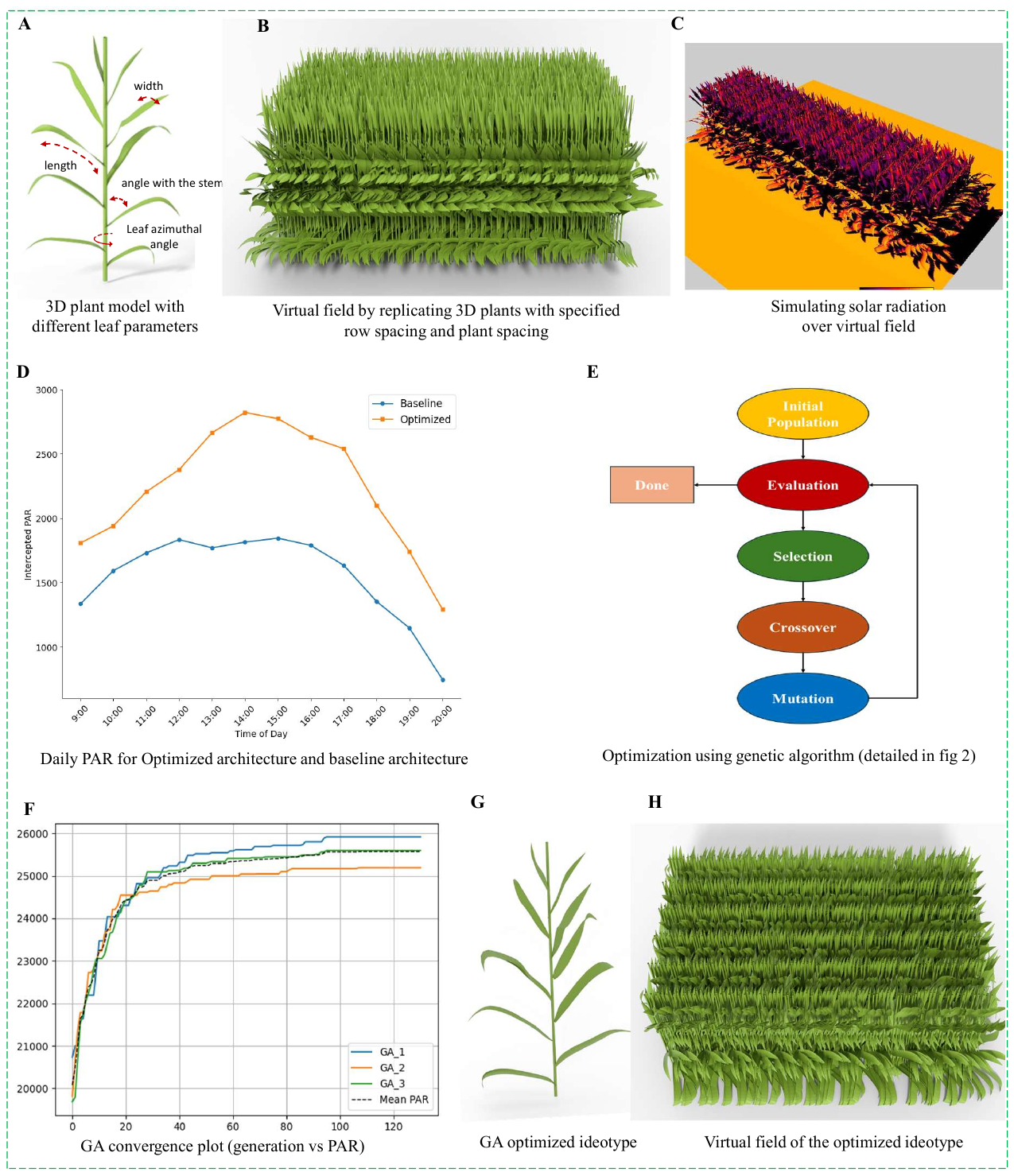}
 \caption{Computational framework for canopy optimization. (A) A parametric 3D maize plant model was constructed with leaf length, width, leaf vertical angle, and azimuthal angle as controllable traits. (B) Virtual fields were generated by replicating these models at fixed row and plant spacing. (C) Reverse ray-tracing simulations quantified diurnal photosynthetically active radiation (PAR) absorption with multiple scattering. (D) Daily intercepted PAR profiles highlight improved interception dynamics compared with the baseline. (E) An evolutionary algorithm (detailed in Fig.~2) iteratively modified plant traits, evaluated PAR interception, and selected beneficial architectures. (F) Convergence plots demonstrate progressive improvement in canopy-level PAR interception across generations. (G) The resulting optimized canopy architecture exhibited enhanced light capture. (H) Virtual field of the optimized ideotype.}
 \label{fig:problem_statement}
\end{figure*}

Traditional breeding, however, relies on incremental selection constrained by the 'breeder's eye' and the logistical bottlenecks of field trials~\cite{hossain2022maize, murchie2022casting}. Consequently, the sheer dimensionality of canopy design—encompassing leaf length, width, vertical, and azimuthal angles—creates a combinatorial landscape too vast for empirical exploration~\cite{murchie2022casting}, leaving potentially beneficial architectural solutions undiscovered. Moreover, many functional-structural plant models~\cite{vos2007functional, vos2010functional} impose fixed allometries or narrow trait ranges, limiting their ability to propose truly novel architectures~\cite{liu2025maize2035, carbajal2024role, he2021modeling, gage2019field}. Here, we present a computational ideotype-design framework that systematically explores this vast trait space and uncovers high-performing canopy architectures beyond the sampling of existing germplasm. Through this process, we identify an emergent ideotype—a coordinated combination of architectural traits—that maximizes canopy-scale light interception under dense planting conditions. Our approach comprises three integrated components:
\begin{itemize}
    \item \textit{Procedural 3D plant modeling:} We build a parameterized representation of maize (fig~\ref{fig:problem_statement}A), wherein each organ (here, leaf) is instantiated with independent trait parameters—length, width, vertical angle (smaller angles indicate more upright leaves), and leaf azimuthal angle. This plant model is then scaled into a virtual field using realistic planting densities (Fig~\ref{fig:problem_statement}B).
    
    \item \textit{Physics-based light interception simulation:} Using a radiative-transfer model~\cite{bailey2018reverse, bailey2019helios}, we compute daily photosynthetically active radiation (PAR)~\cite{mccree1981photosynthetically} intercepted by each virtual canopy under realistic solar trajectories and field-density scenarios~(fig~\ref{fig:problem_statement}E). Our simulation workflow follows a previously documented, validated pipeline; optical parameters and multiple-scattering settings are detailed in the SI and prior work~\cite{saleem2025accessing}.
    \item \textit{Evolutionary algorithm (specifically, genetic algorithm) for optimization:} Evolutionary algorithms systematically improve a set of virtual canopies by recombining and mutating trait vectors, then selecting the highest-performing individuals over many generations. This search heuristic excels in navigating complex, multimodal design landscapes where gradient-based methods falter~\cite{holland1992genetic, whitley1994genetic}. Overview of the framework is shown in Fig~\ref{fig:problem_statement} A-H.
\end{itemize}

By treating key architectural features as free variables, the opimization revealed emergent strategies that \textit{a priori} were not obvious. In addition to the expected vertical stratification of leaf angles consistent with the smart canopy, our optimized ideotype exhibited a distinctive \emph{radially tiled} arrangement of leaves around the stalk: successive leaves were azimuthally offset, reducing self and neighbor-shading \textit{within and between plants}. As detailed in the Results and SI (Sections~S4–S5), the qualitative canopy organization (vertical stratification of angles and widths, tiled azimuths) was preserved across multiple maize-growing latitudes and six spacing regimes, indicating robustness to solar geometry and density.

Beyond uncovering novel trait combinations, our framework offers considerable practical advantages: (i) \textit{Scalability:} We can rapidly evaluate thousands of ideotypes across a range of planting densities, row spacings, and geographic latitudes, enabling region-specific customization; (ii) \textit{Extensibility:} The open-source pipeline can incorporate additional constraints, such as currently accessible allometric constraints, or additional traits like mechanical stability or water-use efficiency, or be adapted to other crop species, or other planting arrangements; and (iii) \textit{Translatability:} Advances in gene-editing technologies like CRISPR/Cas9 now make it feasible to target specific architectural genes independently, surmounting the challenges of genetic linkage and pleiotropy that hinder conventional breeding~\cite{wang2022analysis, wang2022crispr, liu2020high}. In practice, however, implementing these modifications still requires identifying the genes that control each trait and determining how specific edits will produce the desired phenotypic effect. By pinpointing precise trait values, our computational ideotype design can directly inform molecular breeding and genome-editing strategies.

We hypothesize that the emergent computational ideotype will deliver enhanced light-use efficiency under high-density planting, translating to enhanced biomass and grain yields. Ultimately, our procedural ideotype‐design pipeline bridges computational modeling, large-scale optimization, and modern biotechnology to chart a path toward the next generation of maize hybrids. In doing so, it addresses the pressing challenge of sustainably boosting crop productivity to meet escalating global food demands.

% ~\cite{carbajal2024role, christensen2018use, rotter2015use, li2025next}.

\begin{figure*}[t!]
 \centering
\includegraphics[width=\linewidth, height=0.33\textheight, trim=0 150 0 130, clip]{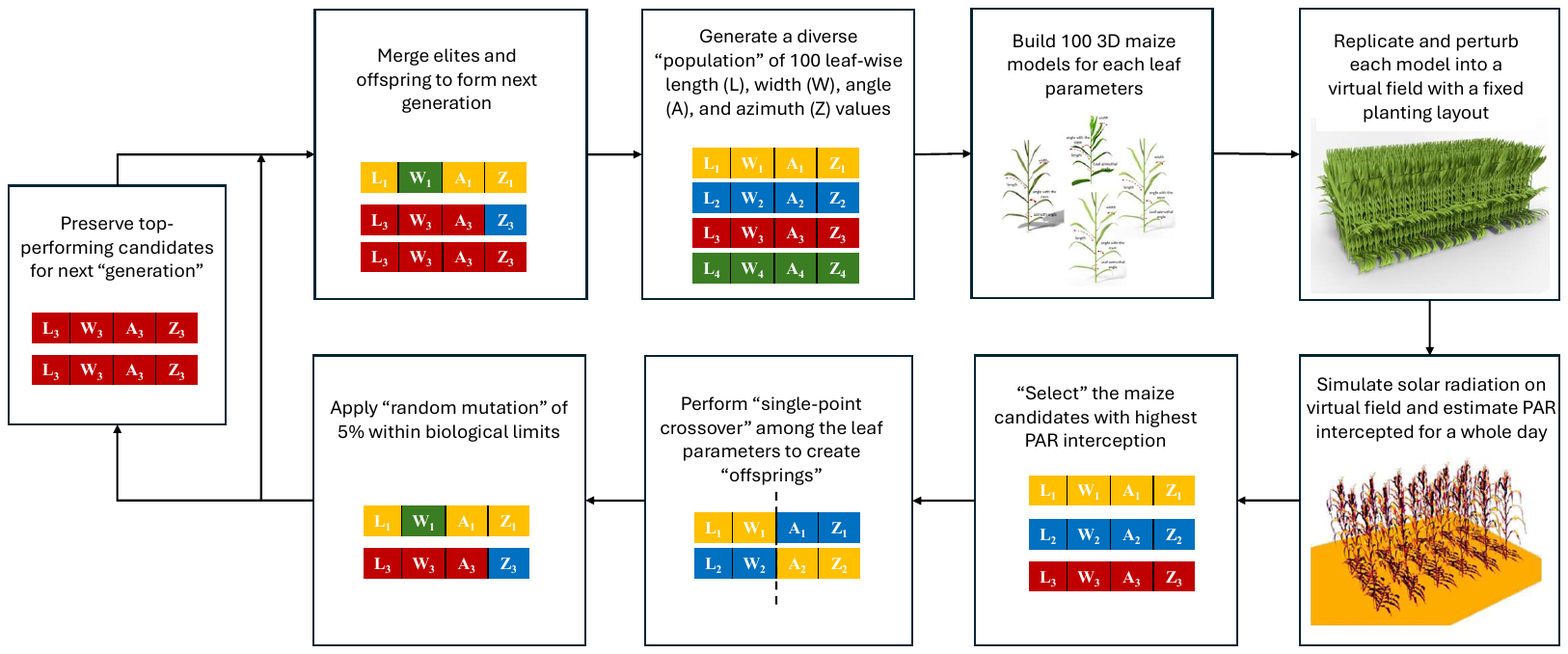}
 \caption{Evolutionary Algorithm framework (specifically, a Genetic Algorithm) for optimizing maize architecture. The process begins with an initial population of maize plant architectures, each defined by key parameters (leaf length, width, vertical angle, and azimuthal angle). The EA iteratively evaluates plant designs based on their light capture efficiency, applies selection, crossover, and mutation operations to generate new candidate solutions, and refines the population over successive generations until an optimal maize architecture emerges.}
\label{fig: GA}
\end{figure*}

\section*{Materials and Methods}

\subsubsection*{Canopy Geometry and Trait Parameterization}
\label{sec:canopy_geometry}

We developed a parametric 3D maize canopy model based on Non-Uniform Rational B-Spline (NURBS) surfaces, incorporating realistic leaf-level structural parameters (hereafter referred to as traits)~\cite{piegl2012nurbs,hadadi2025procedural}. Each virtual plant consisted of a fixed number of leaves (ten per plant, consistent with typical maize architecture) arranged along a central stalk. Each leaf was represented as a NURBS surface, defined by control points that describe its curvature and orientation in 3D space. Four independent traits -- leaf length, maximum width, inclination angle (from the stem), and azimuth (rotation around the vertical axis) -- were parameterized for each leaf. Trait bounds were derived from empirical field measurements across diverse maize genotypes in high-density plots (see Section~S1 in the SI Appendix), and from leaf definitions used in a curated, leaf-annotated 3D maize resource, ensuring that all simulated architectures remained biologically feasible and mesh-interoperable~\cite{MaizeField3D}. These traits capture the dominant structural determinants of canopy light interception while keeping the model computationally efficient for large-scale virtual experiments.

To construct the virtual canopy, a single 3D plant mesh was exported from the NURBS model and replicated in a $6\times6$ grid. Plants were positioned according to a row spacing of 30~inches (0.76~m) and a within-row plant spacing of 6~inches (0.15~m), matching typical high-density planting configurations. To mimic realistic per-plant variability, we did \emph{not} tile identical copies of a candidate: for each plant in the virtual field we independently applied a stochastic azimuthal jitter of $\pm10^{\circ}$ about the vertical axis to capture orientation variability. The individual meshes were combined into a composite field geometry. Periodic boundary conditions were applied to minimize edge effects, effectively simulating an infinite canopy. The resulting field-level mesh was exported as an OBJ file for light interception modeling in Helios~\cite{bailey2019helios}.

Simulations targeted the R1 (silking) stage, approximately 60 days after planting~\cite{abendroth2011corn}. At this stage, the maize canopy architecture is effectively fixed: all leaves are fully expanded, internode elongation has ceased, and leaf angles and orientations remain stable~\cite{abendroth2011corn}. Therefore, we assume a static canopy for modeling light interception. Additionally, from the R1 stage onward, a substantial portion of photosynthates is partitioned toward the kernels, making intercepted PAR a meaningful proxy for yield potential~\cite{tollenaar1999physiology, andrade2002yield, lizaso2017modeling}.

\begin{figure*}[ht]
    \centering
    \includegraphics[width=\linewidth, height=0.35\textheight, trim=0 120 0 130, clip]{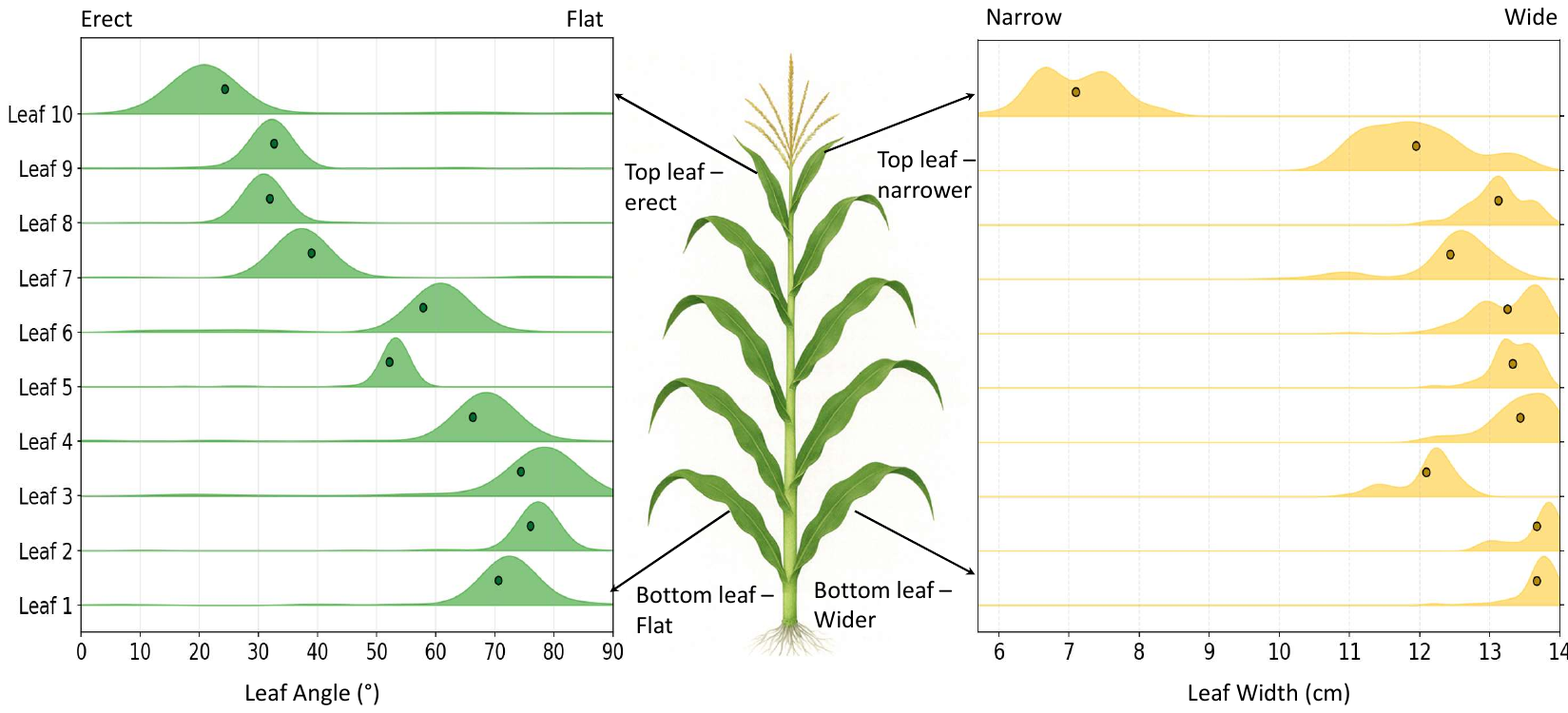}
    \caption{Vertical stratification of leaf morphology in the optimized maize ideotype. The figure shows how leaf angles (left) and widths (right) vary systematically from the upper to lower canopy. Upper leaves are more erect and narrower, whereas lower leaves are flatter and wider. This gradient enhances whole-canopy light distribution by allowing sunlight to penetrate deeper layers while maintaining efficient capture near the top. The density distributions -- plotted using the results from top 10 individuals from 3 distinct EA runs -- illustrate the emergent coordination of angle and width traits identified by the evolutionary optimization algorithm, reflecting a functional architecture similar to field-bred maize hybrids.}
    \label{fig: leaf_values}
\end{figure*}

\subsection*{Radiation Simulation}
\label{sec:radiation_model}
Intercepted PAR was simulated using the radiation module of Helios~\cite{bailey2019helios} (v2.1.0). The model computes dynamic solar position and mutual shading at mesh-level geometric precision. Simulations (for the results in the main text) assumed a representative clear-sky day in Ames, Iowa (42.0$^{\circ}$N, 93.6$^{\circ}$W) on August~7, spanning 07:00–20:00 local time at hourly resolution. Both direct and diffuse components of solar irradiance were modeled, assuming maize leaf reflectance~$=0.05$ and transmittance~$=0.05$ following Earl and Tollenaar~\cite{earl1997maize}. Instantaneous absorbed PAR fluxes (W\,m$^{-2}$) were integrated over the daylight period and converted to daily energy per ground area, expressed as MWh\,acre$^{-1}$\,day$^{-1}$. This daily canopy-integrated absorbed PAR served as the fitness metric for the evolutionary algorithm (EA) optimization described below.  

All Helios configuration parameters, including optical properties and atmospheric inputs, are provided in Section S2 of the SI Appendix. The configuration follows a previously documented, field-referenced pipeline for estimating canopy PAR~\cite{saleem2025accessing}.

\subsection*{Evolutionary Algorithm}

To identify canopy architectures that maximize light capture, we employed an evolutionary algorithm (EA)~\cite{lambora2019genetic}. Each candidate plant was represented as a ``chromosome'' comprising 40 ``gene'', corresponding to four parameters (leaf length, width, inclination, and azimuth) for each of ten leaves. The EA initialized a population of 100 diverse canopy architectures and evolved them over 200 generations to maximize the total daily absorbed photosynthetically active radiation (PAR) per acre. The optimization workflow is summarized in Fig.~\ref{fig: GA}.  

Selection, crossover, and mutation were applied to each generation, promoting high-fitness individuals that intercepted more PAR. The fitness function was defined as the canopy-integrated absorbed PAR across daylight hours (expressed in MWh\,acre$^{-1}$\,day$^{-1}$):
\[
F(\mathbf{x}) = \sum_{t \in \mathcal{T}} A_{\mathrm{PAR}}(\mathbf{x}, t)\,\Delta t,
\]
where $A_{\mathrm{PAR}}(\mathbf{x}, t)$ is the absorbed PAR flux (W\,m$^{-2}$) for canopy configuration $\mathbf{x}$ at time $t$, and $\Delta t$ is the hourly time step. The daily fitness was expressed as MWh\,acre$^{-1}$\,day$^{-1}$.

Three independent EA runs with different random seeds were conducted to confirm reproducibility. The algorithm typically reached convergence after $\sim$80 generations, producing closely related optimized ideotypes. The baseline canopy was parameterized from the field hybrid \textit{B73}, which exhibits moderately erect upper leaves and predominantly perpendicular azimuthal orientations relative to the planting row. All simulations were executed on high-performance computing nodes equipped with NVIDIA A100 GPUs and 32-core CPUs. Each complete optimization (100 individuals $\times$ 200 generations $=$ 20{,}000 canopy evaluations) required  40-48 hours of wall time. Detailed EA hyperparameters, selection operators, and mutation settings are provided in Section~S3 of the SI Appendix.

Building on the computational framework described above, we evaluated how optimization through evolutionary algorithm reshaped canopy architecture and light interception performance under high‐density planting. Simulations were conducted using a parameterized 3D maize model comprising ten leaves per plant, each independently defined by length, width, leaf vertical angle, and azimuthal angle (Fig.~\ref{fig:problem_statement}A). Virtual plants were arranged in a field layout representative of modern high-density conditions, with 30-inch (0.76 m) row spacing and 6-inch (0.15 m) plant spacing, corresponding to approximately $87{,}700~\text{plants}~\text{ha}^{-1}$, and periodic boundary conditions were applied to minimize edge effects. To mimic realistic per‐plant architectual variability, we do \emph{not} tile identical copies of a candidate: instead, for each plant in the virtual field we independently applied a stochastic azimuthal rotation of $\pm10^{\circ}$ about the vertical axis to capture orientation variability.
%instead, for each plant in the virtual field we independently perturbed each leaf trait by $\pm$5\% around the candidate’s values. 
Trait ranges were constrained by field‐measured values from diverse maize genotypes to ensure biological realism~\cite{MaizeField3D} (also see SI for data on ranges). All simulations represented the R1 (silking) stage, when canopy architecture is fixed and photosynthetically active radiation (PAR) interception closely approximates yield potential~\cite{tollenaar1999physiology, lizaso2017modeling}.

\subsubsection*{Optimized Ideotype and Convergence Behavior}
The evolutionary optimization algorithm successfully evolved canopy architectures that substantially enhanced light interception relative to the baseline virtual canopy derived from field‐measured genotypes~\cite{saleem2025accessing}. By iteratively adjusting key morphological traits—including vertical leaf angle, width, length, and azimuthal leaf angle—the optimization explored a wide range of canopy architectures and converged on an ideotype that maximized canopy‐scale PAR interception under dense planting conditions (Fig.~\ref{fig:problem_statement}A).

Across three independent optimization runs with different random initializations, the algorithm consistently converged to a canopy configuration intercepting 56.49 MWh PAR acre$^{-1}$ day$^{-1}$, representing ~13\% improvement over field genotype

with one of the highest PAR interceptions -  48.72 MWh PAR acre$^{-1}$ day$^{-1}$. Fitness trajectories stabilized after approximately 80 generations, indicating convergence of the evolutionary search (Fig.~\ref{fig:problem_statement}F). The similarity among final ideotypes across runs suggests robustness and repeatability of the optimization process. 

The following sub-sections characterize the structural and functional features of these optimized architectures/ideotypes, highlighting how architectural modifications improved light distribution throughout the canopy and their broader implications for breeding and crop design.

\subsubsection*{Light Interception}
Daily PAR interception profiles revealed that optimized canopy architectures consistently outperformed the baseline across all hours of the day (Fig.~\ref{fig:problem_statement}D). Improvements were greatest during  late afternoon when conventional canopies suffer from self‐shading; the optimized architecture maintained higher PAR penetration throughout the canopy.  Simulations across a range of latitudes (SI, Section~S4) showed similar gains, indicating that the ideotype is not over‐fit to a single environment. We further evaluated the optimization across a range of planting densities and row spacings; detailed results are provided in SI, Section~S5. Radiative‐transfer configuration followed a documented ``Helios'' framework (reverse ray tracing with multiple scattering), previously shown to agree with field PAR interception measurements, providing additional confidence in the simulated gains~\cite{bailey2019helios, saleem2025accessing}.
% Integrating PAR intercepted over time yielded the 16\% daily improvement noted above.

\subsubsection*{Trait-Level Changes in Architecture}
The optimized ideotype featured clear vertical stratification of leaf traits. Leaf angles were measured relative to the stalk, such that smaller angles indicate more erect leaves (0$^{\circ}$ = parallel to the stalk, 90$^{\circ}$ = horizontal). Leaves at the top exhibited smaller vertical angles (20$^{\circ}$–50$^{\circ}$), effectively minimizing top-layer shading, whereas lower canopy leaves displayed higher vertical angles (60$^{\circ}$–80$^{\circ}$), improving light interception in shaded regions. A similar vertical gradient was observed in leaf width, where upper leaves were narrower and lower leaves became progressively wider, enabling efficient capture of the residual sunlight that penetrates deeper into the canopy (Fig.~\ref{fig: leaf_values}). This configuration promotes efficient light redistribution by allowing incoming radiation to penetrate deeper into the canopy and be absorbed more uniformly across layers. Such trait stratification emerged consistently across independently evolved solutions, suggesting a convergent optimization strategy driven by canopy-scale light dynamics.

\subsubsection*{Azimuthal Leaf Orientation and Spatial Packing}
We also examined how leaves were azimuthally arranged around the stalk, a trait that influences both self-shading and interplant light competition. Maize naturally exhibits an alternate (distichous) phyllotaxy, with leaves emerging in two opposing ranks along the stem. In field-grown maize, this inherent pattern typically manifests as one of three canopy-level azimuthal orientations (see Fig S2): off-row parallel (perpendicular to the planting row), which minimizes mutual shading; on-row parallel, where leaves align with the row direction; or \textit{random}, where entire individual plants vary in their overall orientation relative to the planting rows~\cite{saleem2025accessing, zhou2024genetic}.

\begin{figure}[ht]
    \centering
    \includegraphics[width=0.65\linewidth, height=0.4\textheight, trim=130 80 135 80, clip]{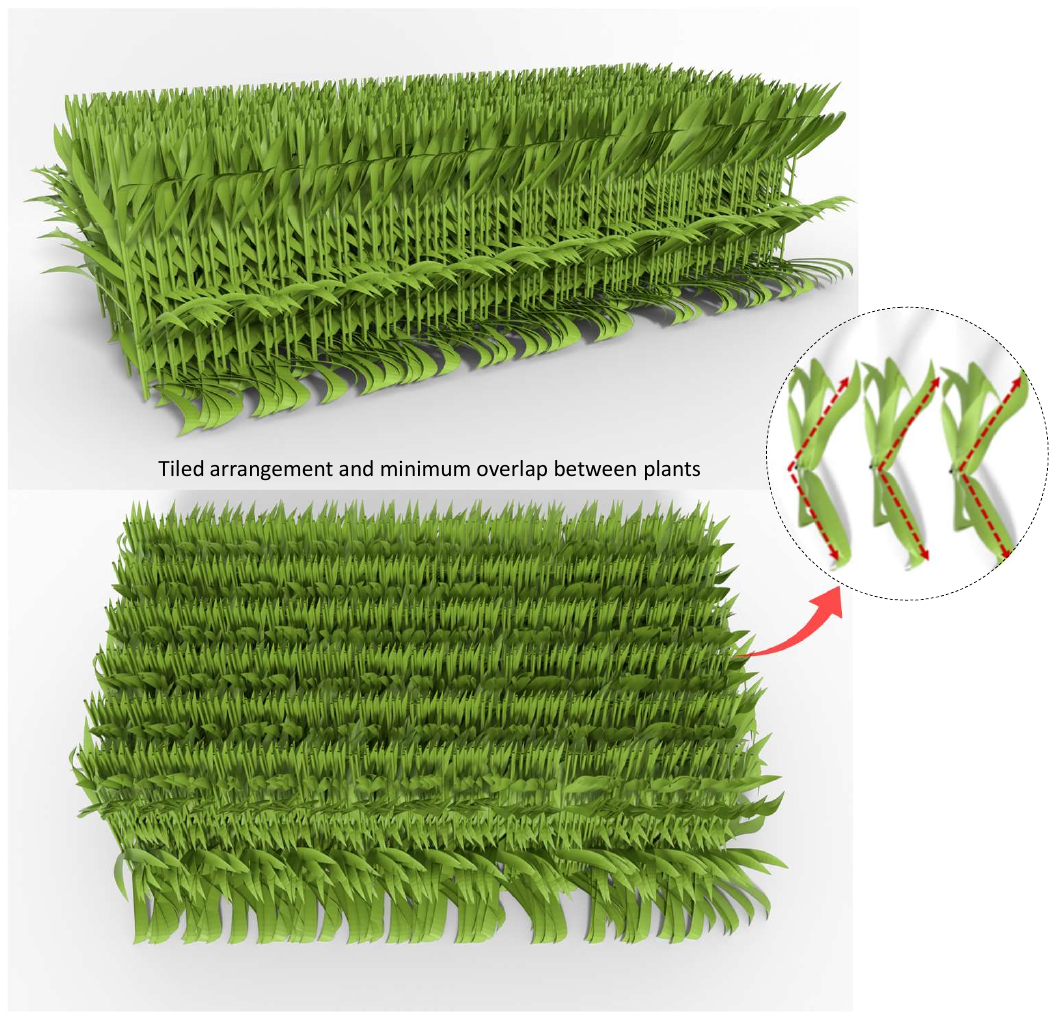}
    \caption{Emergent azimuthal “tiling” in the GA-optimized smart canopy. Oblique (top) and plan (bottom) views of the optimized virtual field show successive leaves azimuthally offset around the stalk (inset), breaking the conventional distichous pattern and reducing within- and between-plant overlap, thereby mitigating mutual shading and improving whole-canopy light interception under dense planting}
    \label{fig: azimuth}
\end{figure}

In contrast to the strict alternate phyllotaxy typical of maize, our EA optimization converged upon a \textit{tiled or quasi-radial arrangement} of leaf azimuths around the stalk (Fig.~\ref{fig: azimuth}). This deviation suggests that while the canonical distichous structure may optimize light capture for isolated plants, it generates excessive mutual shading in dense monocultures. The emergent radial tiling represents a `community-optimal' strategy that minimizes self- and neighbor competition.

%In contrast to this alternate phyllotaxy and its resulting canopy-level orientations, our GA optimization produced a \textit{tiled or quasi-radial arrangement} of leaf azimuths around the stalk (Fig.~\ref{fig: azimuth}), a configuration that differs from the bilateral symmetry exhibited by almost all maize. Although not perfectly uniform, this quasi-radial organization redistributed leaves across multiple azimuths, reducing shading both within and between plants and improving PAR interception under varying solar angles. 
Figure~\ref{fig: azimuth} illustrates this tiled distribution, demonstrating how computational optimization can identify canopy architectures that enhance light-use efficiency and mitigate shading losses.
% \footnote{Many high-performing hybrids exhibit leaves arranged perpendicularly to planting rows, avoiding shading from neighboring plants but potentially increasing self-shading within dense stands.}

\subsubsection*{Similarity to Experimentally Validated F$_1$ and Commercial Hybrids}

The optimized ideotype exhibited architectural features consistent with both experimentally validated and commercially successful maize hybrids, underscoring the practical relevance of computational design. Our results on leaf angles closely resemble the \textit{smart-canopy} architecture characterized by upright upper leaves, moderately inclined middle, and flatter lower leaves—traits that improve light penetration and, consequently, boost grain yield ~\cite{ort2015redesigning, mantilla2017differential, tian2024maize}. The resemblance between our computationally derived canopy and the experimentally validated \textit{lac1}-based F$_1$ hybrids reported by Tian~\textit{et~al.}~\cite{tian2024maize}, which exhibit this smart-canopy configuration, reinforces the physiological credibility and predictive capacity of the optimization framework. A similar pattern occurs in elite commercial hybrids (see, for instance~\citep{PioneerLeafAngle2023}), where erect upper leaves enhance light transmission and density tolerance~\cite{elli2023maize}(Table 1). 
%such as P1151AM{\texttrademark} and P1311AMXT{\texttrademark}

Furthermore, empirical studies reveal that maize can adjust leaf azimuthal orientation in response to shading, often turning leaves toward inter-row spaces to maximize sunlight capture~\citep{zhou2024genetic}. The tiled azimuthal arrangement that emerged in our optimized ideotype mirrors this adaptive strategy, further supporting the model’s ability to reproduce ecophysiologically advantageous behavior. Collectively, these parallels demonstrate that our computational framework captures the core design logic underlying modern high-performance maize, providing a quantitative foundation for accelerating ideotype discovery through simulation-guided breeding.

\subsubsection*{Trait-Level Analyses Confirming Mechanistic Contributions}

To verify that these architectural patterns were mechanistically responsible for improved light capture, we performed trait-restricted optimization experiments in which the EA was restricted to optimize only a subset of parameters while all others were clamped to baseline values. When only leaf angles were free, the algorithm consistently produced ideotypes with \emph{erect upper leaves}, achieving a modest gain in PAR interception (40.72 $\rightarrow$ 43.00 MWh acre$^{-1}$ day$^{-1}$). Optimization of azimuthal angle alone yielded a \emph{tiled leaf orientation} around the stalk, with a comparable improvement (40.72 $\rightarrow$ 42.99 MWh acre$^{-1}$ day$^{-1}$). Allowing angles and widths together led to \emph{erect, narrower upper leaves}, substantially increasing canopy interception (40.72 $\rightarrow$ 49.47 MWh acre$^{-1}$ day$^{-1}$). Finally, when all traits were optimized jointly, the algorithm integrated these features synergistically to reach the highest performance (56.49 MWh acre$^{-1}$ day$^{-1}$).

These trait-restricted analyses demonstrate that the key architectural signatures identified in the full optimization (erect angles, radially tiled azimuths, and thinner upper leaves) emerge independently when traits are optimized in isolation, and that their combination yields additive and synergistic benefits to canopy light capture (Table~\ref{tab:ablation}).

\begin{table}[t]
\centering
\caption{Trait-Level analysis of individual trait contributions to canopy-level light interception. Values represent daily absorbed PAR energy (MWh acre$^{-1}$ day$^{-1}$). The baseline reference genotype intercepted 40.72 MWh acre$^{-1}$ day$^{-1}$.}
\label{tab:ablation}
\begin{tabular}{lc}
\toprule
\textbf{Trait set} & \textbf{PAR} \\
 % & (MWh acre$^{-1}$ d$^{-1}$) & \\
\midrule
Baseline   &  40.72\\ 
Optimize azimuth only       & 42.99 \\
Optimize angles only        & 43.00 \\
Optimize Angles + Width     & 49.47 \\
Optimize all traits (Full)  & 56.49  \\
\bottomrule
\end{tabular}
\end{table}

% \subsubsection*{Extending to Other Environments: Different Locations, and Different Planting Densities}
% The proposed computational framework demonstrates scalability across diverse environmental conditions, enabling ideotype optimization beyond a single maize genotype or climate. By coupling the genetic algorithm with solar radiation simulations, the pipeline can be adapted to optimize plant architecture under varying solar radiation levels, latitudinal gradients, and atmospheric conditions. This adaptability suggests potential for identifying ideotypes suited to distinct climatic zones without requiring extensive field trials.

\subsubsection*{Extending to Other Environments: Different Locations, and Different Planting Densities}

The approach seamlessly extends to diverse environmental and agronomic conditions, enabling the optimization of ideotypes for various planting densities and geographic locations. We conducted independent optimization runs at five distinct latitudes spanning the southmost to northmost limits of commercial maize production above the equator, including three maize-growing sites in the U.S. --- Ames, IA (42$^\circ$N), Thomas County, KS (39$^\circ$N), and Bismarck, ND (47$^\circ$N) --- as well as Barinas, Venezuela (7$^\circ$N) representing the southern extreme and Peace River, Alberta, Canada (56$^\circ$N) representing the northern extreme to assess the impact of solar position on canopy design. While trait magnitudes exhibited modest adjustments (e.g., slightly steeper upper leaves at higher latitudes), the overall architectural patterns, such as vertical stratification of leaf angles and widths, remained consistent (see Section~S4 in SI for additional details). Importantly, this consistency across latitudes indicates that breeders may not need to generate location-specific canopy architectures for commercial production, thereby simplifying the deployment of improved ideotypes at scale.

Additionally, we evaluated the effect of planting density by simulating six representative spacing configurations (various row spacing x seed spacing per row), ranging from $15^" \times 3^"$  to $30^" \times 6^"$. The framework consistently converged to high-performing ideotypes across all spacing regimes, with trait adaptations reflecting local light competition dynamics. For instance, denser spacings yielded more uniform leaf lengths, while wider spacings preserved stronger bottom-to-top length gradients. Despite these variations, the emergent \textit{smart canopy} architecture, characterized by erect upper leaves and wider, flatter lower leaves, was preserved across densities (see Section S5 in the SI for additional details).

These findings demonstrate that the framework can accommodate both environmental heterogeneity and agronomic variability, providing a versatile tool for identifying context-specific ideotypes that complement exhaustive field trials. Beyond maize, this approach can be extended to other crops, provided a detailed 3D parametric model is available. By integrating photosynthesis models, the pipeline could be adapted to optimize additional physiological traits, such as carbon assimilation rates and stomatal conductance, in response to environmental variables. This versatility allows for crop-specific optimization strategies, enhancing productivity and resource-use efficiency in diverse agroecosystems. As functional-structural plant models continue to evolve, coupling evolutionary algorithms with high-resolution environmental simulations presents a powerful avenue for designing climate-resilient crop ideotypes.

\section*{Discussion}
Our results demonstrate that computational optimization of canopy architecture, guided by physically informed light modeling, can uncover both known and previously underappreciated plant configurations that improve light interception. Across independent runs of our evolutionary optimization, we observed consistent emergence of a tiled or radially distributed azimuthal leaf orientation pattern. This architecture contrasts with the alternate leaf orientation common in maize, suggesting a different spatial strategy for minimizing self-shading under dynamic solar angles. Although such spiral/radial azimuthal patterns are rare among maize genotypes, they have been observed in certain lines, such as CM158Q (Fig.~S3), indicating that this configuration is biologically feasible and could potentially be introduced into breeding programs.

By distributing leaves more evenly around the stalk, the spiral/radial azimuthal pattern reduces mutual shading and improves light penetration into deeper canopy layers. The fact that this pattern evolved repeatedly in our EA experiments from randomized starting populations indicates that it is a robust solution under selection for maximizing daily canopy-level PAR.

Beyond azimuth, the optimization favored erect upper leaves and tapered widths—traits known to reduce mutual shading and promote light transmission to lower canopy strata. The preference for upright upper leaves mirrors phenotypes in elite commercial hybrids (e.g., Pioneer P1151AM{\texttrademark}) and aligns with genetic studies on \textit{liguleless1} and \textit{LAC1}, which demonstrate that reduced leaf vertical angle at the canopy top can enhance light penetration and photosynthetic efficiency~\cite{tian2024maize}. These convergences validate the biological realism of the optimization and reinforce empirical evidence that targeted control of leaf angle can improve canopy light-use efficiency. To validate this, future work should involve phenotyping of genetic lines engineered or selected to approximate the optimized trait combinations, followed by field trials under representative agronomic conditions. 

From a breeding perspective, the ability to algorithmically explore and rank thousands of virtual trait combinations provides a powerful complement to traditional field selection. Breeding pipelines are often constrained by long cycle times, environmental variability, and the difficulty of disentangling genetic linkage and pleiotropy. In contrast, computational ideotype design allows breeders to pre-screen architectural configurations under controlled virtual environments, identify promising multi-trait combinations, and focus empirical testing on only the most promising candidates. This targeted approach can reduce trial-and-error in the field, guide marker-assisted selection toward architecture-associated loci, and prioritize gene-editing interventions for traits like leaf angle and azimuthal angle that have clear mechanistic links to light capture.

A major strength of our pipeline is the integration of an evolutionary algorithm with a physics-based reverse ray-tracing model (Helios~\cite{bailey2019helios}) that accounts for dynamic solar position, mutual shading, and multiple scattering with mesh-level geometric precision. This enables evaluation of trait combinations under realistic full-diurnal light conditions, avoiding the oversimplifications common in empirical light-distribution models. By evaluating performance at the R1 stage~\cite{abendroth2011corn}, we target the critical phenological window where canopy architecture is fully established and grain filling begins. This isolates the effects of geometry from growth dynamics, providing a 'static' blueprint that can represent the cumulative architectural goal for breeding programs. Further, the simulation workflow used here follows a previously documented, field-referenced pipeline for estimating canopy PAR~\cite{saleem2025accessing}, and the trait bounds and organ definitions used to build our parametric plants are drawn from segmented, field-derived 3D maize reconstructions with explicit leaf-level parameterization~\cite{MaizeField3D}. Together, these links to empirical data support both the physiological plausibility and reproducibility of the in silico ideotypes.

However, several limitations warrant discussion. 
First, our optimization prioritized PAR interception as the primary driver of yield potential in high-density environments. While we assume well-watered conditions where light is the dominant bottleneck, this single-objective focus allows us to isolate the geometric upper limits of canopy efficiency, establishing a theoretical benchmark against which multi-objective trade-offs (e.g., water-use efficiency, or nutrient distribution) can be evaluated in future work.
%First, our optimization was single-objective, targeting PAR interception without explicitly accounting for potential trade-offs with other physiological processes, such as water-use efficiency, canopy temperature regulation, or nutrient distribution. We assume well-watered conditions in which light is the primary bottleneck. A canopy optimized solely for light capture may not maximize yield if it compromises other resource balances. 
Second, the simulations assume a static architecture at R1, ignoring leaf reorientation, senescence, or seasonal changes in solar elevation. Third, while the EA explores a broad architectural trait space, we do not explicitly enforce any genetic constraints, meaning that some trait combinations may still require complex stacking or decoupling of loci to realize in practice, even if modern gene-editing technologies reduce those barriers. Fourth, although simulations were performed at multiple latitudes to assess the effects of solar geometry, each scenario represented a single clear-sky day rather than a full growing-season light climate.

Despite these constraints, the emergence of biologically plausible, high-performing architectures, many of which align with traits already exploited in elite hybrids, demonstrates the value of this approach for ideotype discovery. The reproducibility of trait emergence across independent EA runs suggests that these solutions are robust rather than artifacts of stochastic search. More broadly, the pipeline establishes a path toward simulation-guided ideotype design: using empirically grounded 3D plant parameterizations to generate testable, high-value canopy blueprints before committing to multi-year breeding cycles. Even so, validating a GA-generated ideotype in the field would still require multiple years of breeding and validation.

\section*{Conclusion}
This study introduces an end-to-end framework for computational smart canopy ideotype design, combining evolutionary algorithm, mechanistic 3D plant modeling, and physics-based light simulation to identify maize architectures that maximize canopy-level light interception. The approach recovered known advantageous traits (e.g., erect upper leaves) and revealed underappreciated configurations (e.g., quasi-radial azimuthal arrangements) that challenge conventional assumptions about optimal plant form. Results were robust across latitudes and planting densities (Section S4 and S5 in SI), and the pipeline is grounded in empirically derived organ-level geometry and a validated radiative-transfer configuration, strengthening confidence in the simulated gains.

By decoupling the search for architectural solutions from the constraints of conventional breeding, this method enables exploration of ideotype spaces that are otherwise inaccessible. As gene-editing technologies (e.g., CRISPR/Cas9) make precise modulation of individual traits feasible, the pipeline provides a principled basis for prioritizing targets and trait combinations. For breeders, such computational screening functions serve as a virtual testbed, identifying high-value phenotypes before committing resources to multi-year, multi-location trials. This accelerates the breeding cycle and increases the probability of field success by focusing efforts on trait combinations with clear mechanistic advantages.

Future extensions should expand the framework to optimize full-season ideotypes by simulating canopy performance over the entire growing cycle, rather than a single day or phenological stage (here R1). Incorporating temporal dynamics such as changing solar elevation, phenology, and leaf senescence would provide a more nuanced evaluation of architectural performance. In parallel, integrating mechanistic photosynthesis (coupled with CO$_2$, temperature, humidity, and wind) would enable the direct estimation of carbon gain and water-use efficiency, thereby bridging the gap between light interception and physiological productivity. Embedding these biophysical models within multi-objective optimization and accelerating evaluation with machine-learned surrogate models would support scalable exploration of genotype–environment–management interactions across climates.

This work highlights the promise of algorithmic plant design for sustainably increasing crop productivity. By marrying biological realism with computational exploration, ideotype discovery becomes an engineered, first-principles process by complementing, rather than replacing, field innovation to deliver architectures tailored for dense cultivation and a changing climate.

\subsection*{Data and Software Availability}
All source codes, input configurations, and analysis scripts are available upon request. Full parameter files for Helios radiation simulations and EA runs are provided in the SI Appendix.

\subsection*{Acknowledgements}
We acknowledge funding from the AI Research Institutes program supported by NSF and USDA-NIFA under AI Institute: for Resilient Agriculture, Award No. 2021-67021-35329, from NSF 2412929, and from Plant Science Institute at ISU.

% \bibsplit[11]
%Use \bibsplit to split the references from the body of the text. Value "[2]" represents the number of reference in the left column (Note: Please avoid single column figures & tables on this page.)

\bibliographystyle{elsarticle-num-names} 
\bibliography{references}

@article{erenstein2022global,
  title={Global maize production, consumption and trade: trends and R\&D implications},
  author={Erenstein, Olaf and Jaleta, Moti and Sonder, Kai and Mottaleb, Khondoker and Prasanna, Boddupalli M},
  journal={Food security},
  volume={14},
  number={5},
  pages={1295--1319},
  year={2022},
  publisher={Springer}
}

@article{hadadi2025procedural,
  title={Procedural generation of 3D maize plant architecture from LiDAR data},
  author={Hadadi, Mozhgan and Saraeian, Mehdi and Godbersen, Jackson and Jubery, Talukder Z and Li, Yawei and Attigala, Lakshmi and Balu, Aditya and Sarkar, Soumik and Schnable, Patrick S and Krishnamurthy, Adarsh and others},
  journal={Computers and Electronics in Agriculture},
  volume={236},
  pages={110382},
  year={2025},
  publisher={Elsevier}
}

@article{MaizeField3D,
  title={AgriField3D: A Curated 3D Point Cloud and Procedural Model Dataset of Field-Grown Maize from a Diversity Panel},
  author={Kimara, Elvis and Hadadi, Mozhgan and Godbersen, Jackson and Balu, Aditya and Jubery, Talukder and Li, Yawei and Krishnamurthy, Adarsh and Schnable, Patrick S and Ganapathysubramanian, Baskar},
  journal={arXiv preprint arXiv:2503.07813},
  year={2025}
}

@article{nyirenda2021delving,
  title={Delving into possible missing links for attainment of food security in Central Malawi: Farmers’ perceptions and long term dynamics in maize (Zea mays L.) production},
  author={Nyirenda, Harrington and Mwangomba, Wantwa and Nyirenda, Ellen M},
  journal={Heliyon},
  volume={7},
  number={5},
  year={2021},
  publisher={Elsevier}
}

@article{mansfield2014survey,
  title={Survey of plant density tolerance in US maize germplasm},
  author={Mansfield, Brian D and Mumm, Rita H},
  journal={Crop Science},
  volume={54},
  number={1},
  pages={157--173},
  year={2014},
  publisher={Wiley Online Library}
}

@article{assefa2018analysis,
  title={Analysis of long term study indicates both agronomic optimal plant density and increase maize yield per plant contributed to yield gain},
  author={Assefa, Yared and Carter, Paul and Hinds, Mark and Bhalla, Gaurav and Schon, Ryan and Jeschke, Mark and Paszkiewicz, Steve and Smith, Stephen and Ciampitti, Ignacio A},
  journal={Scientific reports},
  volume={8},
  number={1},
  pages={4937},
  year={2018},
  publisher={Nature Publishing Group UK London}
}

@article{yan2024photosynthetic,
  title={Photosynthetic capacity and assimilate transport of the lower canopy influence maize yield under high planting density},
  author={Yan, Yanyan and Duan, Fengying and Li, Xia and Zhao, Rulang and Hou, Peng and Zhao, Ming and Li, Shaokun and Wang, Yonghong and Dai, Tingbo and Zhou, Wenbin},
  journal={Plant Physiology},
  volume={195},
  number={4},
  pages={2652--2667},
  year={2024},
  publisher={Oxford University Press US}
}

@article{kalogeropoulos2024historical,
  title={Historical increases of maize leaf area index in the US Corn Belt due primarily to plant density increases},
  author={Kalogeropoulos, George and Elli, Elvis F and Trifunovic, Slobodan and Archontoulis, Sotirios V},
  journal={Field Crops Research},
  volume={318},
  pages={109615},
  year={2024},
  publisher={Elsevier}
}

@article{saleem2025accessing,
  title={Accessing the Effect of Phyllotaxy and Planting Density on Light Interception in Field-Grown Maize using 3D Reconstructions},
  author={Saleem, Nasla and Jubery, Talukder Zaki and Balu, Aditya and Zhou, Yan and Li, Yawei and Schnable, Patrick S and Krishnamurthy, Adarsh and Ganapathysubramanian, Baskar},
  journal={Smart Agricultural Technology},
  pages={101566},
  year={2025},
  publisher={Elsevier}
}

@article{donald1968breeding,
  title={The breeding of crop ideotypes},
  author={Donald, Colin Malcolm},
  journal={Euphytica},
  volume={17},
  number={3},
  pages={385--403},
  year={1968},
  publisher={Springer}
}

@article{mock1975ideotype,
  title={An ideotype of maize},
  author={Mock, JJ and Pearce, RB},
  journal={Euphytica},
  volume={24},
  number={3},
  pages={613--623},
  year={1975},
  publisher={Springer}
}

@article{carbajal2024role,
  title={The role of the ideotype in future agricultural production},
  author={Carbajal-Friedrich, Alonso AJ and Burgess, Alexandra J},
  journal={Frontiers in Plant Physiology},
  volume={2},
  pages={1341617},
  year={2024},
  publisher={Frontiers Media SA}
}

@article{gauffreteau2018using,
  title={Using ideotypes to support selection and recommendation of varieties},
  author={Gauffreteau, Arnaud},
  journal={OCL},
  volume={25},
  number={6},
  pages={D602},
  year={2018},
  publisher={EDP Sciences}
}

@article{tian2024maize,
  title={Maize smart-canopy architecture enhances yield at high densities},
  author={Tian, Jinge and Wang, Chenglong and Chen, Fengyi and Qin, Wenchao and Yang, Hong and Zhao, Sihang and Xia, Jinliang and Du, Xian and Zhu, Yifan and Wu, Lishuan and others},
  journal={Nature},
  volume={632},
  number={8025},
  pages={576--584},
  year={2024},
  publisher={Nature Publishing Group UK London}
}

@article{ort2015redesigning,
  title={Redesigning photosynthesis to sustainably meet global food and bioenergy demand},
  author={Ort, Donald R and Merchant, Sabeeha S and Alric, Jean and Barkan, Alice and Blankenship, Robert E and Bock, Ralph and Croce, Roberta and Hanson, Maureen R and Hibberd, Julian M and Long, Stephen P and others},
  journal={Proceedings of the national academy of sciences},
  volume={112},
  number={28},
  pages={8529--8536},
  year={2015},
  publisher={National Academy of Sciences}
}

@article{duvick2005contribution,
  title={The contribution of breeding to yield advances in maize (Zea mays L.)},
  author={Duvick, Donald N},
  journal={Advances in agronomy},
  volume={86},
  pages={83--145},
  year={2005},
  publisher={Elsevier}
}

@article{han2024molecular,
  title={Molecular mechanisms underlying coordinated responses of plants to shade and environmental stresses},
  author={Han, Run and Ma, Liang and Terzaghi, William and Guo, Yan and Li, Jigang},
  journal={The Plant Journal},
  volume={117},
  number={6},
  pages={1893--1913},
  year={2024},
  publisher={Wiley Online Library}
}

@article{sedgley1991appraisal,
  title={An appraisal of the Donald ideotype after 21 years},
  author={Sedgley, RH},
  journal={Field Crops Research},
  volume={26},
  number={2},
  pages={93--112},
  year={1991},
  publisher={Elsevier}
}

@article{murchie2022casting,
  title={Casting light on the architecture of crop yield},
  author={Murchie, Erik H and Burgess, Alexandra J},
  journal={Crop and environment},
  volume={1},
  number={1},
  pages={74--85},
  year={2022},
  publisher={Elsevier}
}

@incollection{hossain2022maize,
  title={Maize breeding},
  author={Hossain, Firoz and Muthusamy, Vignesh and Bhat, Jayant S and Zunjare, Rajkumar U and Kumar, Santosh and Prakash, Nitish R and Mehta, Brijesh K},
  booktitle={Fundamentals of field crop breeding},
  pages={221--258},
  year={2022},
  publisher={Springer}
}

@article{he2021modeling,
  title={Modeling maize canopy morphology in response to increased plant density},
  author={He, Liang and Sun, Weiwei and Chen, Xiang and Han, Liqi and Li, Jincai and Ma, Yuanshan and Song, Youhong},
  journal={Frontiers in Plant Science},
  volume={11},
  pages={533514},
  year={2021},
  publisher={Frontiers Media SA}
}

@article{gage2019field,
  title={In-field whole-plant maize architecture characterized by subcanopy rovers and latent space phenotyping},
  author={Gage, Joseph L and Richards, Elliot and Lepak, Nicholas and Kaczmar, Nicholas and Soman, Chinmay and Chowdhary, Girish and Gore, Michael A and Buckler, Edward S},
  journal={The Plant Phenome Journal},
  volume={2},
  number={1},
  pages={1--11},
  year={2019},
  publisher={Wiley Online Library}
}

@article{liu2025maize2035,
  title={Maize2035: A decadal vision for intelligent maize breeding},
  author={Liu, Hai-Jun and Liu, Jie and Zhai, Zhiwen and Dai, Mingqiu and Tian, Feng and Wu, Yongrui and Tang, Jihua and Lu, Yanli and Wang, Haiyang and Jackson, David and others},
  journal={Molecular Plant},
  year={2025},
  publisher={Elsevier}
}

@article{vos2010functional,
  title={Functional--structural plant modelling: a new versatile tool in crop science},
  author={Vos, J and Evers, Jochem B and Buck-Sorlin, G Hh and Andrieu, Bruno and Chelle, Micha{\"e}l and De Visser, Pieter HB},
  journal={Journal of experimental Botany},
  volume={61},
  number={8},
  pages={2101--2115},
  year={2010},
  publisher={Oxford University Press}
}

@article{vos2007functional,
  title={Functional-structural plant modelling in crop production: adding a dimension},
  author={Vos, Jan and Marcelis, LFM and Evers, JB},
  journal={Frontis},
  pages={1--12},
  year={2007}
}

@Article{bailey2018reverse,
  author    = {Bailey, Brian N},
  journal   = {Ecological Modelling},
  title     = {A reverse ray-tracing method for modelling the net radiative flux in leaf-resolving plant canopy simulations},
  year      = {2018},
  pages     = {233--245},
  volume    = {368},
  publisher = {Elsevier},
}

@Article{bailey2019helios,
  author    = {Bailey, Brian N},
  journal   = {Frontiers in Plant Science},
  title     = {{Helios:} {A} scalable {3D} plant and environmental biophysical modeling framework},
  year      = {2019},
  pages     = {1185},
  volume    = {10},
  publisher = {Frontiers Media SA},
}

@incollection{mccree1981photosynthetically,
  title={Photosynthetically active radiation},
  author={McCree, Keith J},
  booktitle={Physiological plant ecology I: responses to the physical environment},
  pages={41--55},
  year={1981},
  publisher={Springer}
}

@inproceedings{lambora2019genetic,
  title={Genetic algorithm-A literature review},
  author={Lambora, Annu and Gupta, Kunal and Chopra, Kriti},
  booktitle={2019 international conference on machine learning, big data, cloud and parallel computing (COMITCon)},
  pages={380--384},
  year={2019},
  organization={IEEE}
}

@article{wang2022crispr,
  title={CRISPR-Cas technology opens a new era for the creation of novel maize germplasms},
  author={Wang, Youhua and Tang, Qiaoling and Pu, Li and Zhang, Haiwen and Li, Xinhai},
  journal={Frontiers in Plant Science},
  volume={13},
  pages={1049803},
  year={2022},
  publisher={Frontiers Media SA}
}

@article{wang2022analysis,
  title={Analysis of the utilization and prospects of CRISPR-Cas technology in the annotation of gene function and creation new germplasm in maize based on patent data},
  author={Wang, Youhua and Tang, Qiaoling and Kang, Yuli and Wang, Xujing and Zhang, Haiwen and Li, Xinhai},
  journal={Cells},
  volume={11},
  number={21},
  pages={3471},
  year={2022},
  publisher={MDPI}
}

@article{liu2020high,
  title={High-throughput CRISPR/Cas9 mutagenesis streamlines trait gene identification in maize},
  author={Liu, Hai-Jun and Jian, Liumei and Xu, Jieting and Zhang, Qinghua and Zhang, Maolin and Jin, Minliang and Peng, Yong and Yan, Jiali and Han, Baozhu and Liu, Jie and others},
  journal={The Plant Cell},
  volume={32},
  number={5},
  pages={1397--1413},
  year={2020},
  publisher={American Society of Plant Biologists}
}

@article{jiang2025leaf,
  title={Leaf angle regulation toward a maize smart canopy},
  author={Jiang, Qinyue and Wang, Yijun},
  journal={The Plant Journal},
  volume={121},
  number={2},
  pages={e17208},
  year={2025},
  publisher={Wiley Online Library}
}

@article{elli2023maize,
  title={Maize leaf angle genetic gain is slowing down in the last decades},
  author={Elli, Elvis F and Edwards, Jode and Yu, Jianming and Trifunovic, Slobodan and Eudy, Douglas M and Kosola, Kevin R and Schnable, Patrick S and Lamkey, Kendall R and Archontoulis, Sotirios V},
  journal={Crop science},
  volume={63},
  number={6},
  pages={3520--3533},
  year={2023},
  publisher={Wiley Online Library}
}

@article{pepper1977leaf,
  title={Leaf orientation and yield of maize 1},
  author={Pepper, GE and Pearce, RB and Mock, JJ},
  journal={Crop Science},
  volume={17},
  number={6},
  pages={883--886},
  year={1977},
  publisher={Wiley Online Library}
}

@article{whitley1994genetic,
  title={A genetic algorithm tutorial},
  author={Whitley, Darrell},
  journal={Statistics and computing},
  volume={4},
  number={2},
  pages={65--85},
  year={1994},
  publisher={Springer}
}

@article{holland1992genetic,
  title={Genetic algorithms},
  author={Holland, John H},
  journal={Scientific american},
  volume={267},
  number={1},
  pages={66--73},
  year={1992},
  publisher={JSTOR}
}

@article{abendroth2011corn,
  title={Corn growth and development},
  author={Abendroth, Lori J and Elmore, Roger Wesley and Boyer, Matthew J and Marlay, Stephanie and others},
  year={2011},
  publisher={Iowa State University Ames}
}

@incollection{tollenaar1999physiology,
  title={Physiology of maize},
  author={Tollenaar, M and Dwyer, LM},
  booktitle={Crop yield: physiology and processes},
  pages={169--204},
  year={1999},
  publisher={Springer}
}

@article{andrade2002yield,
  title={Yield responses to narrow rows depend on increased radiation interception},
  author={Andrade, Fernando H and Calvino, Pablo and Cirilo, Alfredo and Barbieri, Pablo},
  journal={Agronomy Journal},
  volume={94},
  number={5},
  pages={975--980},
  year={2002},
  publisher={Wiley Online Library}
}

@article{lizaso2017modeling,
  title={Modeling the response of maize phenology, kernel set, and yield components to heat stress and heat shock with CSM-IXIM},
  author={Lizaso, Jon I and Ruiz-Ramos, Margarita and Rodr{\'\i}guez, Luc{\'\i}a and Gabaldon-Leal, Clara and Oliveira, JA and Lorite, Ignacio J and Rodr{\'\i}guez, Alfredo and Maddonni, Gustavo Angel and Otegui, Maria Elena},
  journal={Field Crops Research},
  volume={214},
  pages={239--254},
  year={2017},
  publisher={Elsevier}
}

@misc{PioneerLeafAngle2023,
  author       = {{Pioneer Hi-Bred International, Inc.}},
  title        = {Corn Leaf Angle Response to Plant Density},
  year         = {2023},
  howpublished = {\url{https://www.pioneer.com/us/agronomy/corn-leaf-angle-response-plant-density.html}},
  note         = {Accessed: 2025-08-06}
}

@article{zhou2024genetic,
  title={Genetic regulation of self-organizing azimuthal canopy orientations and their impacts on light interception in maize},
  author={Zhou, Yan and Kusmec, Aaron and Schnable, Patrick S},
  journal={The Plant Cell},
  volume={36},
  number={5},
  pages={1600--1621},
  year={2024},
  publisher={Oxford University Press US}
}

@article{earl1997maize,
  title={Maize leaf absorptance of photosynthetically active radiation and its estimation using a chlorophyll meter},
  author={Earl, HJ and Tollenaar, M},
  journal={Crop Science},
  volume={37},
  number={2},
  pages={436--440},
  year={1997},
  publisher={Wiley Online Library}
}

@book{piegl2012nurbs,
  title={The NURBS book},
  author={Piegl, Les and Tiller, Wayne},
  year={2012},
  publisher={Springer Science \& Business Media}
}

\end{document}